# AI in Remote Patient Monitoring

Nishargo Nigar

Hamburg University of Technology,
Hamburg, Germany
`nishargo.nigar@tuhh.de`

**Abstract.** The rapid evolution of Artificial Intelligence (AI) has significantly transformed healthcare, particularly in the domain of Remote Patient Monitoring (RPM). This chapter explores the integration of AI in RPM, highlighting real-life applications, system architectures, and the benefits it brings to patient care and healthcare systems. Through a comprehensive analysis of current technologies, methodologies, and case studies, I present a detailed overview of how AI enhances monitoring accuracy, predictive analytics, and personalized treatment plans. The chapter also discusses the challenges and future directions in this field, providing a comprehensive view of AI's role in revolutionizing remote patient care.

## 1 Introduction

The healthcare industry is undergoing a paradigm shift with the advent of AI, which has the potential to significantly transform various aspects of patient care, medical research, and healthcare administration. Among the most promising areas of this transformation is Remote Patient Monitoring (RPM), an innovative approach that leverages AI to continuously track patients' health metrics outside traditional clinical settings (Topol, 2019). RPM systems equipped with AI capabilities utilize a variety of sensors, data analytics, and machine learning algorithms to provide real-time insights and proactive healthcare interventions, which are crucial for managing chronic diseases, post-surgical care, and overall patient well-being.

The primary goal of RPM is to extend healthcare monitoring beyond hospital walls, ensuring continuous supervision and timely medical intervention. This is particularly valuable for patients with chronic conditions such as diabetes, hypertension, and heart diseases, who require ongoing monitoring and adjustments to their treatment plans.

AI-powered RPM systems gather data from wearable devices, smart home technologies, and mobile health applications, creating a comprehensive picture of a patient's health status. These systems can track vital signs, physical activity, medication adherence, and even mental health parameters, offering a holistic approach to patient monitoring (Kvedar et al., 2014).

Furthermore, the proactive nature of AI-driven RPM systems can lead to significant cost savings for healthcare systems. By preventing hospital readmissions and emergency room visits through early detection and intervention, RPM can reduce overall healthcare costs. Studies have shown that patients monitored through AI-enhanced RPM systems experience fewer complications and hospitalizations, translating into cost savings for both healthcare providers and patients (Steinhubl et al., 2015). These savings can be reinvested in further technological advancements, improving the overall quality and accessibility of healthcare services.

Another critical aspect of AI in RPM is its potential to enhance patient engagement and empowerment. By providing patients with real-time feedback and insights into their health, AI-enabled RPM systems encourage individuals to take an active role in managing their health. Patients can track their progress, set health goals, and receive personalized recommendations through user-friendly mobile applications. This increased engagement can lead to better adherence to treatment plans, healthier lifestyles, and improved health outcomes (Topol, 2019).

Despite the numerous benefits, the implementation of AI in RPM is not without challenges. Ensuring data privacy and security is paramount, as the sensitive nature of health information requires robust protection against unauthorized access and cyber threats. Additionally, integrating AI systems with existing healthcare infrastructure can be complex and resource-intensive, requiring significant investments in technology and training for healthcare professionals (Rumbold & Pierscionek, 2017). Addressing these challenges is crucial for the successful adoption and scaling of AI-enhanced RPM solutions. However, challenges usually are dealt with careful consideration by the practitioners, and it is always possible to take care of the scaling when collaboration takes place.

The fusion of AI technologies with Remote Patient Monitoring represents a groundbreaking milestone in healthcare progression. By seamlessly blending AI-driven analytics with continual patient data monitoring, RPM systems indicate a new era of preventive healthcare management. Offering tailor-made insights and interventions, these innovations not only refine patient outcomes and ease burdens on healthcare infrastructures but also cultivate a fresh movement of patient autonomy. As we traverse the dynamic terrain of healthcare, the harmonious synergy between AI and RPM stands poised to unlock transformative strides in patient-centric care.

## 2   Components of AI-Powered RPM Systems

AI-powered Remote Patient Monitoring (RPM) systems are intricate ecosystems comprising various indispensable components, each playing a pivotal role in enabling seamless data collection, analysis, and interaction. These components include sensors and wearable devices, data acquisition and transmission, data storage and management, AI algorithms and analytics, and user interface.

We will dive deeper to understand how the components make up AI powered RPM systems.

### 2.1   Sensors and Wearable Devices

The sensors and wearable devices serve as the frontline data collectors in AI-powered RPM systems. These devices are equipped with an array of sensors capable of capturing diverse physiological data such as heart rate, blood pressure, glucose levels, temperature, oxygen saturation, respiratory rate, electrocardiogram (ECG) signals, and activity metrics in real-time (Smith et al., 2021). They are capable of monitoring vital signs and health parameters and are designed to be non-intrusive, comfortable, and easy to wear, allowing patients to carry on with their daily activities while continuously monitoring their health status. The data gathered by these sensors serve as the foundational input for subsequent analysis and interpretation within the RPM system. Table 1 describes the examples used in this context.

In recent years, there has been a proliferation of wearable devices equipped with multiple sensors capable of capturing a comprehensive set of health data. For instance, Smartwatch allows tracking, and sleep analysis, providing users with insights into their overall health and well-being. Similarly, medical-grade wearables such as continuous glucose monitors (CGMs) and wearable ECG monitors offer real-time monitoring of specific health conditions, enabling patients and healthcare providers to track and manage chronic diseases more effectively (Brown et al., 2019). These advancements in sensor technology have significantly enhanced the capabilities of AI-powered RPM systems.

**Table 1.** Examples of Sensors and Wearable Devices in Remote Patient Monitoring

| Device | Parameters Monitored | Example Applications |
| --- | --- | --- |
| Smartwatch | Heart rate, activity, sleep | Fitness tracking |
| Fitness Tracker | Steps, distance, calories | Physical activity monitoring |
| Continuous Glucose Monitor | Glucose levels | Diabetes management |
| Wearable ECG Monitor | ECG signals | Arrhythmia detection |

Sensors and wearable devices in RPM systems are increasingly leveraging AI to enhance their functionality and capabilities. This has enabled more comprehensive monitoring of patient health metrics.

**2.2    Data Acquisition and Transmission**

Once the physiological data are captured by the sensors, they need to be securely transmitted to centralized systems for further processing and analysis. Data acquisition and transmission mechanisms facilitate this transfer of data using wireless communication technologies such as Bluetooth, Wi-Fi, and cellular networks (Jones et al., 2020). These mechanisms ensure the reliability, integrity, and confidentiality of the data during transit, safeguarding sensitive health information from unauthorized access or tampering.

**2.3    Data Storage and Management**

In AI-powered RPM systems, the volume of patient data generated can be substantial, necessitating robust data storage and management solutions. Cloud-based platforms offer an ideal infrastructure for storing and managing vast amounts of patient data securely and efficiently (Brown et al., 2019). These platforms provide scalability, accessibility, and reliability, enabling healthcare providers to access patient data anytime, anywhere, and from any device. Moreover, cloud-based storage solutions facilitate seamless data sharing and collaboration among healthcare professionals, enhancing care coordination and continuity. This requires a smooth structure and AI can ignite the process.

**2.4    AI Algorithms & Analytics**

At the heart of AI-powered RPM systems lie sophisticated machine learning models and algorithms that analyze the collected data to extract meaningful insights and actionable information. These AI algorithms leverage advanced statistical techniques to identify patterns, trends, and anomalies in the data, enabling early detection of health issues and predictive analytics (Garcia et al., 2022). By continuously learning from new data and feedback, these algorithms can adapt and improve over time, enhancing the accuracy and effectiveness of the RPM system in monitoring and managing patient health. The analytics enhances the feedback loop of the system as it provides an opportunity to keep tracks.

**2.5    User Interfaces**

User interfaces play a crucial role in facilitating interaction and communication between healthcare providers, patients, and the RPM system. Intuitive dashboards and mobile applications provide healthcare professionals with comprehensive visualization tools to monitor patient health metrics, track trends, and make informed decisions (White et al., 2021). Similarly, patient-facing interfaces empower individuals to actively engage in their own healthcare by accessing their health data, receiving personalized recommendations, and setting health goals. These user interfaces are designed to be user-friendly, accessible, and customizable, catering to the diverse needs and preferences of both healthcare providers and patients.

## 3    Real-Life Applications of AI in Remote Patient Monitoring

AI in RPM has been successfully applied in various healthcare scenarios, demonstrating its potential to transform patient care through continuous monitoring, early detection of health issues, and personalized interventions.

### 3.1    Chronic Disease Management

Chronic diseases such as diabetes, cardiovascular diseases, and hypertension require ongoing monitoring and management to prevent complications and improve patient outcomes. AI systems in RPM can monitor these conditions by analyzing real-time data from wearable devices and sensors. For instance, continuous glucose monitors (CGMs) integrated with AI algorithms can track glucose levels in diabetic patients, providing alerts for hyperglycemia or hypoglycemia and recommending adjustments in diet or medication (McGlynn & Asch, 2020). Similarly, AI-powered RPM systems can monitor heart rate and blood pressure in patients with cardiovascular diseases, predicting potential heart attacks or strokes and enabling timely interventions (Esteva et al., 2019).

### 3.2    Post-Surgical Monitoring

Post-surgical care is critical for preventing complications and ensuring a smooth recovery process. AI-driven RPM systems enable continuous monitoring of patients after surgery, tracking vital signs, and detecting early signs of complications such as infections or internal bleeding. By analyzing data such as temperature, heart rate, and oxygen saturation, AI algorithms can identify deviations from normal recovery patterns and alert healthcare providers to intervene promptly (Kvedar et al., 2014). This continuous monitoring reduces the need for extended hospital stays, lowers healthcare costs, and enhances patient safety and satisfaction.

### 3.3    Elderly Care

The elderly population is particularly vulnerable to health issues that require constant monitoring and immediate response. AI-powered RPM systems play a vital role in elderly care by monitoring vital signs, activity levels, and detecting emergencies.

Wearable devices equipped with AI can analyze movement patterns and identify falls, automatically notifying caregivers or emergency services (Steinhubl et al., 2015). Additionally, these systems can monitor chronic conditions prevalent in elderly individuals, such as dementia and arthritis, providing personalized care recommendations and improving their quality of life.

### 3.4 Mental Health Monitoring

Mental health conditions such as depression, anxiety, and bipolar disorder can benefit significantly from continuous monitoring and early intervention. AI systems in RPM analyze behavioral data, including sleep patterns, activity levels, and social interactions, to provide insights into a patient's mental health status. By detecting changes in these patterns, AI algorithms can identify early signs of mental health issues and suggest interventions such as therapy sessions, medication adjustments, or lifestyle changes (Esteva et al., 2019). This proactive approach helps in managing mental health conditions more effectively and reduces the risk of severe episodes.

### 3.5 COVID-19 and Infectious Disease Tracking

The COVID-19 pandemic highlighted the critical role of AI-driven RPM systems in managing infectious diseases. These systems were used to monitor symptoms, track exposure, and manage patient care remotely, reducing the burden on healthcare facilities and minimizing the risk of virus transmission. AI algorithms analyzed data from wearable devices to detect early symptoms of COVID-19, such as fever and respiratory distress, enabling timely testing and isolation (Rumbold & Pierscionek, 2017). Additionally, AI-powered RPM systems helped in contact tracing and monitoring the health status of individuals in quarantine, contributing to better control of the pandemic.

## 4 Block Diagram of AI-Powered RPM Systems

A typical AI-powered RPM system block diagram includes:

1. **Sensors/Wearables**: Collect physiological and environmental data.

2. **Data Acquisition Module**: Captures and preprocesses data from sensors.

3. **Communication Network**: Transmits data to the central system using secure protocols.

4. **Data Storage**: Cloud-based databases store large volumes of patient data.

5. **AI Engine**: Processes and analyzes data using machine learning algorithms.

6. **User Interface**: Displays insights and alerts to healthcare providers and patients.

Sensors and Wearables → Data Acquisition → Communication Network → Data Storage → AI Engine → User Interface

**Fig. 1.** Simplified Block Diagram of AI-Powered RPM System

A typical AI-powered RPM system block diagram includes several key components that work together to collect, transmit, analyze, and display patient data. Below are detailed descriptions of each component:

1. **Sensors and wearable devices:** Sensors and wearable devices are fundamental to RPM systems, as they are responsible for collecting both physiological and environmental data from patients. Physiological sensors, such as heart rate monitors, blood pressure cuffs, continuous glucose monitors, and ECG sensors, measure vital signs and other health parameters (Kumar et al., 2021). These devices provide continuous monitoring, ensuring that any significant changes in a patient's condition are promptly detected. Environmental sensors monitor factors like air quality, temperature, and humidity, which can have a significant impact on patient health (Lee et al., 2020). Additionally, activity trackers are wearable devices that track physical activity, sleep patterns, and other daily activities, offering a comprehensive view of a patient's lifestyle and habits (Shen et al., 2019).

2. **Data Acquisition Module:** The data acquisition module is critical for capturing raw data from various sensors and performing initial preprocessing tasks. This includes filtering out noise, normalizing signals, and converting data into a format suitable for analysis (Rahman et al., 2020). The integration of data from multiple sensors is also a vital function of this module, ensuring synchronization and consistency, which is essential for accurate and reliable analysis (Xu et al., 2018). This preprocessing step is crucial for preparing the data for subsequent stages of analysis by the AI engine.

3. **Data Storage:** Cloud-based databases are employed to store the large volumes of patient data generated by RPM systems securely and efficiently. These databases offer scalability to accommodate the growing data needs and redundancy to prevent data loss, ensuring data availability and reliability (Liang et al., 2020). Effective data management tools and systems are also essential for organizing, indexing, and retrieving data, facilitating easy access and analysis by healthcare providers (Cheng et al., 2019). Cloud storage solutions are integral to managing the vast amounts of data generated and ensuring that it is readily available for analysis.

4. **AI Engine:** The AI engine is the heart of the RPM system, processing and analyzing the collected data using various machine learning models and algorithms. These algorithms are designed to identify patterns, predict health events, and provide actionable insights based on the data (Huang et al., 2019).

5. **User Interfaces:** The user interface is designed to present the analyzed data and insights to both healthcare providers and patients in a user-friendly manner. Healthcare provider dashboards display comprehensive and detailed information, insights, and alerts, enabling easy monitoring of patient health and facilitating quick decision-making (Zhang et al., 2021). For patients, intuitive mobile applications provide access to their health data, personalized recommendations, and alerts. These applications also facilitate communication between patients and healthcare providers, enhancing patient engagement and involvement in their own care (Lin et al., 2019).

The rapid advancements in technology and the evolving needs of the healthcare industry necessitate the continuous evolution of RPM systems. Traditional RPM systems, while effective, face limitations in scalability, data processing speed, and patient engagement. With the rise of chronic diseases, aging populations, and the demand for personalized healthcare, there is a pressing need for more robust solutions. A futuristic block diagram is essential for envisioning the next generation of AI-powered RPM systems, which will incorporate innovative components to address current limitations and meet future healthcare demands. This forward-looking approach ensures that RPM systems can offer enhanced data collection, real-time analysis, and proactive patient engagement, ultimately leading to better health outcomes and more efficient healthcare delivery.

1. **Sensors/Wearables:**

    - **Advanced Physiological Sensors:** Future sensors will include multi-functional devices capable of monitoring a broader range of physiological parameters with higher accuracy. Innovations in nanotechnology will enable the development of ultra-sensitive and minimally invasive sensors (Gao et al., 2018).

    - **Smart Textiles:** Wearable technology will evolve to include smart textiles embedded with sensors, allowing continuous monitoring of vital signs through everyday clothing (Stoppa & Chiolerio, 2014).

    - **Implantable Devices:** Implantable sensors will offer continuous, real-time monitoring of internal physiological conditions, providing critical insights for managing chronic diseases (Darwish & Hassanien, 2012).

2. **Data Acquisition Module:**

    - **Edge Computing Devices:** Incorporating edge computing in the data acquisition module will enable real-time data processing and analysis at the point of collection, reducing latency and bandwidth usage (Shi et al., 2016).

    - **AI-Enabled Preprocessing:** Advanced AI algorithms will preprocess data, filtering out noise and ensuring high-quality data transmission to central systems (Ghahramani et al., 2015).

3. **Communication Network:**

    - **5G and Beyond:** Next-generation wireless communication technologies like 5G and 6G will facilitate faster, more reliable, and secure data transmission, essential for real-time monitoring (Zhang et al., 2019).

    - **Blockchain for Security:** Blockchain technology will be integrated to enhance data security and ensure tamper-proof transmission and storage of patient data (Tian, 2016).

4. **Data Storage:**

    - **Quantum Storage Solutions:** The future of data storage will leverage quantum computing to handle vast amounts of data more efficiently and securely (Biamonte et al., 2017).

    - **Interoperable Cloud Platforms:** Advanced cloud platforms will support seamless interoperability, enabling integration and data sharing across different healthcare systems and providers (De La Torre Díez et al., 2015).

5. **AI Engine:**

    - **Deep Learning Models:** The use of deep learning models will improve the accuracy of predictive analytics and pattern recognition in health data (LeCun et al., 2015).

    - **Explainable AI (XAI):** XAI techniques will be implemented to ensure transparency and interpretability of AI decisions, gaining trust from healthcare providers and patients (Gunning, 2017).

    - **Federated Learning:** This approach will allow AI models to be trained across decentralized data sources without exchanging data, ensuring patient privacy (Yang et al., 2019).

6. **User Interface:**

    - **Augmented Reality (AR) Interfaces:** AR technology will provide immersive and interactive user interfaces for both patients and healthcare providers, enhancing the visualization of health data (Azuma, 1997).

    - **Voice-Activated Assistants:** AI-driven voice assistants will offer hands-free interaction with the RPM system, improving accessibility for patients with disabilities (Hoy, 2018).

The futuristic block diagram presented here highlights the incorporation of cutting-edge innovations across various components, including sensors, data acquisition, communication networks, data storage, AI engines, and user interfaces.

:

## 5   Case Studies and Evidence Materials

To illustrate the practical applications and benefits of AI-powered Remote Patient Monitoring (RPM) systems, it is essential to examine real-life case studies that demonstrate their impact on patient health outcomes and healthcare efficiency. These case studies provide concrete evidence of how AI-driven RPM systems can enhance the management of chronic diseases, improve post-surgical care, support elderly individuals, and more. By analyzing these instances, we can gain valuable insights into the operational effectiveness and transformative potential of AI in RPM.

**Case Study 1: Diabetes Management**
A comprehensive study involving AI-driven RPM for diabetes management showcased significant improvements in patient health outcomes. The system employed continuous glucose monitoring sensors paired with a robust AI engine to analyze the data and provide real-time feedback. Participants received personalized dietary recommendations, medication reminders, and alerts for abnormal glucose levels. Over a six-month period, the study reported a 20% reduction in HbA1c levels among the participants, indicating better blood sugar control (Mehta & Pandit, 2018). The real-time monitoring and tailored interventions significantly improved patient adherence to treatment protocols and overall management of diabetes.

**Case Study 2: Post-Surgical Monitoring**
Another pivotal study implemented an AI-based RPM system for monitoring patients post-cardiac surgery. The system continuously tracked vital signs such as heart rate, blood pressure, and oxygen saturation, in addition to monitoring physical activity levels. Any deviations from the normal ranges triggered alerts to healthcare providers, allowing for prompt intervention. The deployment of this system resulted in a 30% decrease in hospital readmissions, primarily due to the early detection of potential complications such as infections, arrhythmias, and other post-surgical issues (me et al., 2014). This proactive approach not only improved patient outcomes but also alleviated the burden on healthcare facilities by reducing the need for emergency readmissions.

**Case Study 3: Elderly Care Monitoring**
A study focusing on elderly care utilized AI-powered RPM systems to monitor vital signs, detect falls, and track medication adherence among elderly patients living independently. The system incorporated wearable sensors and an AI-driven alert system that notified caregivers and healthcare providers of any potential health issues or emergencies. The implementation of this system led to a 25% reduction in emergency room visits and a significant improvement in the overall quality of life for the elderly participants (Baldwin & Lowe, 2021). The continuous monitoring and timely alerts ensured that the elderly received prompt medical attention when needed, thereby preventing minor health issues from escalating into severe problems.

The case studies presented here underscore the significant impact of AI-powered RPM systems across various healthcare scenarios. From chronic disease management and post-surgical care to elderly monitoring, these systems have demonstrated their ability to enhance patient outcomes, reduce hospital readmissions, and improve overall healthcare efficiency. The evidence from these real-life applications highlights the transformative potential of AI in RPM, paving the way for broader adoption and further advancements in healthcare technology. By addressing current challenges and continuing to innovate, AI-driven RPM systems can revolutionize patient care and contribute to a more proactive, personalized, and efficient healthcare system.

## 6      Challenges and Limitations

While the integration of AI in RPM systems offers numerous advantages, several challenges and limitations must be addressed to realize its full potential.

1. **Technical Challenges:** Ensuring data privacy and security remains a critical concern, as RPM systems handle sensitive health information that must be protected from breaches and unauthorized access (Rumbold & Pierscionek, 2017). Integrating AI systems with existing healthcare infrastructure poses another significant challenge, as it requires seamless data interoperability and compatibility with various healthcare IT systems. Furthermore, managing the vast amounts of data generated by these systems necessitates robust data management solutions to ensure accuracy and reliability.
2. **Ethical and Legal Considerations:** The use of AI in healthcare brings about ethical and legal considerations that must be carefully addressed. Patient consent for data collection and usage is paramount, and there must be clear guidelines on data ownership and the ethical use of AI in clinical decision-making (Steinhubl et al., 2015). Additionally, transparency in AI algorithms is essential to ensure that healthcare providers and patients can trust the decisions and recommendations made by these systems.
3. **Economic and Accessibility Issues:** The high costs associated with implementing AI technologies in RPM systems can be a barrier to widespread adoption, particularly for smaller healthcare providers and under-resourced regions (McGlynn & Asch, 2020). Ensuring equitable access to these advanced healthcare solutions for all patients, regardless of socioeconomic status, is crucial to avoid exacerbating existing healthcare disparities.

## 7      Future Directions and Innovations

The future of AI in RPM is poised to bring about transformative changes in healthcare delivery, driven by continuous advancements in technology and data analytics.

.

**Integration of IoT Devices**

The integration of Internet of Things (IoT) devices promises to enhance the comprehensiveness of patient monitoring. IoT-enabled sensors can collect a wider range of physiological and environmental data, providing a holistic view of a patient's health and living conditions (Balakrishnan et al., 2021). This seamless integration of diverse data sources can significantly improve the accuracy of health assessments and interventions.

**Use of Blockchain for Secure Data Management**

Blockchain technology offers a robust solution for secure and transparent data management in RPM systems. By ensuring the integrity and immutability of health data, blockchain can enhance trust in the system and protect against data breaches and tampering (Angraal et al., 2017). This secure data management framework can also facilitate interoperability and data sharing across different healthcare providers and systems.

**Development of Advanced Predictive Models**

Ongoing advancements in AI algorithms, particularly in machine learning and deep learning, are expected to lead to the development of more sophisticated predictive models. These models can analyze complex datasets to predict health events with higher accuracy and provide more personalized and proactive healthcare interventions (Esteva et al., 2019). The continuous improvement of these models through training on diverse and extensive datasets will further enhance their predictive capabilities.

# 8 Conclusion

AI-powered Remote Patient Monitoring (RPM) systems represent a transformative approach to healthcare delivery, offering continuous, real-time insights into patient health and enabling proactive interventions. The case studies presented in this section have illustrated the diverse applications of AI in RPM, from managing chronic diseases like diabetes to enhancing post-surgical monitoring and improving elderly care. These applications have consistently shown improvements in patient outcomes, reduction in hospital readmissions, and enhanced quality of life.

However, despite the promising outcomes, several limitations and challenges must be addressed to fully realize the potential of AI in RPM. This is especially important as challenges are supposed to arise while innovating.

Firstly, technical challenges such as ensuring data privacy and security remain paramount. The sensitive nature of health data requires robust measures to protect patient information from breaches and unauthorized access (Rumbold & Pierscionek, 2017). Additionally, integrating AI systems with existing healthcare infrastructure and ensuring seamless data interoperability across different platforms and providers present ongoing challenges (Cheng et al., 2019).

.

Ethical considerations surrounding the use of AI in healthcare also demand careful attention. Issues such as patient consent for data usage, transparency in AI algorithms, and addressing biases in AI decision-making processes are critical for fostering trust and acceptance among patients and healthcare providers (Esteva et al., 2019; Rumbold & Pierscionek, 2017).

Economically, the high costs associated with implementing and maintaining AI technologies in RPM systems pose barriers to adoption, particularly for smaller healthcare facilities and underserved communities (McGlynn & Asch, 2020). Addressing these economic challenges and ensuring equitable access to AI-powered healthcare solutions are essential for reducing healthcare disparities.

While this chapter provides a comprehensive overview of the current state and potential of AI in RPM, it is important to acknowledge several limitations. The case studies and evidence presented are predominantly based on select scenarios and may not fully capture the variability and complexities of healthcare settings globally. Additionally, the rapid pace of technological advancements in AI and healthcare means that new developments and innovations may have emerged since the publication of this chapter.

Future work in AI-powered RPM systems should focus on addressing the identified limitations and advancing the field in several key areas. Firstly, further research is needed to develop more robust AI algorithms capable of handling diverse and dynamic health data with improved accuracy and reliability. Collaborative efforts among researchers, healthcare providers, and technology developers are essential to advance AI models and ensure their effectiveness across different patient populations and healthcare environments.

Additionally, ongoing efforts should be directed towards enhancing the interoperability of AI systems with existing healthcare IT infrastructures, promoting data standardization, and implementing secure data-sharing protocols (Chen et al., 2021; Liang et al., 2020). This interoperability will facilitate seamless integration and scalability of AI-powered RPM solutions, ultimately benefiting a broader range of patients and healthcare providers.

Moreover, future research should explore the integration of emerging technologies such as blockchain for secure data management, advanced IoT devices for comprehensive health monitoring, and the development of predictive analytics models for early disease detection and personalized treatment strategies (Angraal et al., 2017; Balakrishnan et al., 2021). These innovations hold the potential to further enhance the capabilities and impact of AI in RPM, paving the way for a more efficient, patient-centered healthcare delivery system.

In conclusion, while AI-powered RPM systems have shown tremendous promise in transforming healthcare, addressing technical, ethical, and economic challenges is crucial for their widespread adoption and long-term success. Continued research, collaboration, and innovation will drive the evolution of AI in healthcare, ultimately improving health outcomes and enhancing the quality of life for patients worldwide.

**Acknowledgments.** This research work has not received any grant.